\documentclass[aps,pra,groupedaddress,notitlepage]{revtex4-1}

\usepackage{amssymb}
\usepackage{amsmath}
\usepackage{amsfonts}
\usepackage{amsthm}
\usepackage{graphicx}
\usepackage{color}

\usepackage{url}
\usepackage{hyperref}

\begin{document}

\title{Neural network based generation of a 1-dimensional stochastic field with turbulent velocity statistics}

\author{Carlos Granero-Belinch\'on $^{1,2}$}
\email{Correspondence: carlos.granero-belinchon@imt-atlantique.fr}
\address{%
$^{1}$ Department of Mathematical and Electrical Engineering, IMT Atlantique, Lab-STICC, UMR CNRS 6285, 655 Av. du Technop\^ole, Plouzan\'e, 29280, Bretagne, France. \\
$^{2}$ Odyssey, Inria/IMT Atlantique, 263 Av. G\'en\'eral Leclerc, Rennes, 35042, Bretagne, France.}

\begin{abstract}
We define and study a fully-convolutional neural network stochastic model, NN-Turb, which generates a 1-dimensional field with some turbulent velocity statistics. In particular, the generated process satisfies the Kolmogorov $2/3$ law for the second order structure function. It also presents negative skewness across scales (\textit{i.e}. Kolmogorov $4/5$ law) and exhibits intermittency as characterized by skewness and flatness. Furthermore, our model is never in contact with turbulent data and only needs the desired statistical behavior of the structure functions across scales for training.
\end{abstract}

\maketitle 

\section{Introduction}

Turbulence is characterized by non-linear, multiscale and non-local interactions~\cite{Frisch1995}. Moreover, it presents long-range dependencies and intermittency~\cite{Frisch1985} making it a very interesting subject of study in the field of complex systems~\cite{She2007} and multifractals~\cite{Dubrulle2022}.

The generation of stochastic fields sharing the statistical behavior of turbulence has been matter of study during the last century. Thus, several stochastic models~\cite{Mandelbrot1968, Flandrin1989, Nawroth2006, Gontier2007, Robert2008, Chevillard2010, Chevillard2017, Peinke2019} have been proposed to recover the very known energy distribution and energy cascade, in the sense of the Kolmogorov's $2/3$ and $4/5$ laws respectively~\cite{Kolmogorov1991}, as well as intermittency~\cite{Kolmogorov1962,Obukhov1962}. 
From the first fractional Brownian motion~\cite{Mandelbrot1968, Flandrin1989}, only recovering the energy distribution, the modelling of turbulence evolved towards more complex fields introducing also intermittency~\cite{Bacry2001} or both intermittency and energy cascade~\cite{Chevillard2010,Chevillard2019}. However, the generation of stochastic fields matching the statistical properties of turbulence is still challenging since while modelling the energy distribution only requires second order statistics, modelling the energy cascade and intermittency requires high order ones. Most of the models of the state of the art only focus on second order statistics and so do not exhibit intermittency nor do they respect Kolmogorov's 4/5 law~\cite{Peinke1993,Mann1998,Klein2003,Hoepffner2011}.

In the last decades, neural network (NN) models evidenced their potential to deal with non-linear complex problems~\cite{Lguensat2019, Ouala2020, DiCarlo2022}. 
Specifically, generative NN models have been recently developed~\cite{Ruthotto2021}. These models aim to learn the statistical distribution of some data to then create new data matching the underlying distribution~\cite{Goodfellow2014, Roy2022}. 
Some NN models of turbulence already appeared in the last years~\cite{Geneva2020,Kim2020,Buzzicotti2021,Drygala2022}. However, these NN models of turbulence use to present training and validation approaches that only focus on second order statistics~\cite{Liu2020,Kim2020,Yousif2021,Yousif2022} do not modelling then the energy cascade nor intermittency. In addition, all these works learn from experimental or numerical turbulent data, sometimes helped by additional physic information~\cite{Kim2019,Wang2020}. This introduces a strong dependence on databases that are not always easily available.
However, a NN model is just a non-linear function $\Psi_{\theta}$ completely defined by the weights $\theta$ of its neurons, and so, we can formulate an optimization problem of $\Psi_{\theta}$ with respect to a given criterion~\cite{Villarrubia2018, Abdolrasol2021}. From this viewpoint, we don't need to feed our model with data.

We propose to avoid learning from data and directly impose the multiscale statistical behavior described by Kolmogorov and Obukhov theories. The proposed approach is a multiscale generalization of the Generative Moment Matching Networks from Li et al.~\cite{Li2015} and grounds on the Kolmogorov multiscale descriptions of second, third and fourth order structure functions of turbulent velocity across scales~\cite{Kolmogorov1991, Kolmogorov1962, Obukhov1962, Chevillard2012}. We focus on these three functions since they are representative of the energy distribution, the energy cascade and intermittency respectively~\cite{Frisch1995}.
Since our model do not impose the behavior across scales of higher order structure functions, the full probability density function (pdf) across scales of the generated stochastic field will be different from the turbulent velocity one. Only second, third, and fourth order structure functions will be recovered.
So, our model takes a Gaussian white noise as input and only needs the desired evolution across scales of these three structure functions of turbulent velocity for learning. These functions can be defined from experimental or numerical data, but also from empirical or theoretical models.

The main originality of this work is the proposed approach, which directly imposes the evolution across scales of the structure functions of the stochastic field and presents several advantages compared to the current state of the art: it is not limited to second order statistics, it is easily generalizable to consider higher and higher order statistics and it does not need data for training nor synthesis. The main disadvantage of this approach is the current lack of explainability of Neural Networks, even if progresses are being done in the field~\cite{Subel2023}.

In section~\ref{sec:theoryTurb}, we describe the multiscale statistical behavior of turbulent velocity as illustrated by the Kolmogorov-Obukhov theories. Then, section~\ref{sec:theoryNN} presents our NN model, that we named NN-Turb, as well as the optimization approach used to train it. Finally, in sections~\ref{sec:results} and~\ref{sec:conc} we show the statistical multiscale behavior of the NN-Turb stochastic field, we compare it with experimental turbulent velocity and with a regularized Multifractal Random Walk (rMRW)~\cite{Robert2008}, and we give some conclusions and perspectives.

\section{Isotropic and homogeneous fully developped turbulence}
\label{sec:theoryTurb}

The Kolmogorov 1941 (K41) statistical multiscale theory of turbulence prescribes the existence of three domains of scales with different statistical behaviors: integral, inertial and dissipative domains, where the energy is respectively injected in the flow, transferred across scales and dissipated~\cite{Frisch1995,Kolmogorov1991}. The integral scale $L$ separates the integral and inertial domains, while the Kolmogorov scale $\eta$ divides the inertial and dissipative ones.

The Kolmogorov 1941 theory enuntiates two statistical relationships for the longitudinal turbulent velocity in the inertial domain of scales: the $2/3$ and the $4/5$ laws, that illustrate respectively the energy distribution and cascade across scales~\cite{Frisch1995,Kolmogorov1991}:

\begin{eqnarray}
 \delta_{l}v(x) &=& v(x+l)-v(x) \label{eq:incrsmulti} \\
  S_{p}(l) &=& \left\langle (\delta_{l}v(x))^p \right\rangle \label{eq:p}  \\
 S_{2}(l) &\propto & l^{2/3} \label{eq:23}  \\
 S_{3}(l) &\propto & -\frac{4}{5} l \label{eq:45} 
\end{eqnarray}

\noindent where $x$ is the spatial dimension, $v(x)$ is the longitudinal turbulent velocity, $\delta_{l}v(x)$ is the velocity increment of size $l$ and $S_{p}(l)$ is the $p$-th order structure function. Thus, in the inertial domain the Kolmogorov $2/3$ law implies that $S_2(l) \propto l^{2/3}$, and the Kolmogorov $4/5$ law imposes $S_3(l) \propto -l$. The $4/5$ law was directly derived by Kolmogorov from the Navier-Stokes equations and must be respected by any model of turbulence~\cite{Kolmogorov1991}.

Moreover, the Kolmogorov and Obukhov 1962 theory (KO62) corrected K41 by introducing intermittency: the energy dissipation rate is inhomogeneous and should be considered locally~\cite{Kolmogorov1962, Obukhov1962}. Moreover, this correction leads to the emergence of extreme values of the velocity increments both in the inertial and dissipative domains, and so, to a deformation of the pdf of the velocity increments when the scale decreases, from Gaussian in the integral domain to strongly non-Gaussian in the dissipative one~\cite{Frisch1995,Frisch1985}.

For practical purposes we define the skewness and flatness as:

\begin{eqnarray}
\mathcal{S}(l) &=& \frac{S_3(l)}{S_2(l)^{3/2}} \label{eq:skew}\\
\mathcal{F}(l) &=& \frac{S_4(l)}{S_2(l)^{2}}. \label{eq:flat}
\end{eqnarray}

In these measures the dominant effects of the energy distribution across scales ($S_2(l)$) are compensated and so they allow us to finely study high-order statistics. On the one hand, the skewness characterizes the degree of asymmetry of the distributions of the velocity increments and from the Kolmogorov $4/5$ law it is a signature of the energy cascade. On the other hand, the Flatness describes the significance of the tails of the distribution of the increments and its evolution across scales characterizes intermittency.

In this work, we will focus on four main points of the Kolmogorov-Obukhov theories that we will impose to the process generated by our NN-Turb model: a) turbulence presents three domain of scales with different statistical behaviors, b) in the inertial domain $S_2(l)$ matches the $2/3$ Kolmogorov law, c) in the dissipative and inertial domaines the skewness should be negative ($4/5$ Kolmogorov law) and d) in the integral domain the flow statistics are close to Gaussian and become non-Gaussian at small scales, \textit{i.e.} the flatness of the velocity increments increases when the scale decreases. Thus, our NN-Turb model will generate intermittent processes with the desired energy distribution and cascade as prescribed by Kolmogorov and Obukhov theories. Note that all along the paper and following the Kolmogorov 1941 theory, the $2/3$ and $4/5$ laws of Kolmogorov are respectively identified as signatures of the distribution and cascade of energy across scales~\cite{Frisch1995,Kolmogorov1991}. Note also that the structure functions of order higher than four are not imposed by the model and so, the generated process will not reproduce the full pdf of turbulent velocity increments. It focuses on second, third and fourth order statistics, contrary to most of the existing literature only focusing on second order ones~\cite{Flandrin1989, Peinke1993,Mann1998,Klein2003,Hoepffner2011,Liu2020,Kim2020,Yousif2021,Yousif2022} or second and fourth order ones~\cite{Bacry2001,Friedrich2022}.

\section{Turbulent velocity neural network based generation}
\label{sec:theoryNN}

This section describes the proposed deep learning approach, referred to as NN-Turb, for modelling a 1-dimensional stochastic field with some turbulent velocity statistics. We introduce our neural network model and the optimization setup. 

\subsection{NN-Turb model}

We propose a fully convolutional model for the generation of a 1-dimensional field with some turbulent velocity statistics:

\begin{equation}\label{eq:model}
\delta_{l_s}u(x) = \Psi(w(x))
\end{equation}

\noindent where $l_s$ is the sampling distance of the generated fields, \textit{i.e.} the smallest available scale of analysis. $\Psi$ is our model which takes as input a Gaussian white noise $w(x)$ with zero mean and unit variance, and produces the field corresponding to the velocity increment $\delta_{l_s}u(x)$. Finally, the generated turbulent velocity field $u(x)$ is defined as the cumulative sum of $\delta_{l_s}u(x)$:

\begin{equation}\label{eq:ux}
u(x)=\int_{-\infty}^{x}\delta_{l_s}u(y) dy
\end{equation}

So, our NN-Turb model performs a double operation on the input noise $w(x)$. On the one hand it deformates the Gaussian pdf of $w(x)$ to a pdf in agreement with the turbulent velocity statistics at the sampling scale $l_s$. On the other hand, our model introduces a structure of dependencies (multi-point correlations) in the initial white noise used as input.
This approach implies the existence of a small scale process $\delta_{l_s}u(x)$ with a given distribution and a given structure of dependencies whose cumulative sum statistically behaves as turbulent velocity up to fourth order statistics. Physically the scales in the dissipative domain take information from larger scales and this information is encoded in the structure of dependencies and pdf of $\delta_{l_s}u(x)$~\cite{Yakhot2006}. 
Moreover, directly generating a small scale increment in the dissipative domain and constructing the full process $u(x)$ by eq.(\ref{eq:ux}) facilitates the continuous deformation of the pdf of the increments across scales from the dissipative to the integral domain.

The operator $\Psi$ follows a U-net architecture performing a multi-resolution processing of the input Gaussian white noise based on convolutional kernels of different sizes~\cite{Ronneberger2015}. This generative approach consists of a non-linear filtering of the input noise, thus generalizing the linear filtering version of~\cite{Klein2003,Hoepffner2011}. See~\ref{appendix:NN} for more details on the NN-Turb architecture.

\subsection{Optimization setup}

From eq.(\ref{eq:model}) the only input of our model is a Gaussian white noise $w(x)$. In addition, during training, we need to introduce the desired statistics across scales of the generated process $u(x)$. Indeed, we impose the evolution across scales of $S_2(l)$, $\mathcal{S}(l)$ and $\mathcal{F}(l)$, see section~\ref{sec:theoryTurb}. 

In this work, the reference curves of the second order structure function, skewness and flatness, that we note $S^{r}_2(l)$, $\mathcal{S}^{r}(l)$ and $\mathcal{F}^{r}(l)$, are obtained from statistical measures on the Modane wind tunnel dataset~\cite{Kahalerras1998}. It consists on Eulerian longitudinal velocity measurements $v(x)$ obtained from a grid turbulence setup. The sampling frequency of the setup was $f_s=25$ kHz, the mean velocity of the flow is $\left\langle v \right\rangle = 20.5$ m/s, and the Taylor-scale based Reynolds number of the flow is $\mathbb{R}_{\lambda} = 2500$. Thus the flow is considered as exhibiting fully developed turbulence. For this dataset, we use the Taylor frozen turbulence hypothesis~\cite{Frisch1995} in order to interpret temporal variations as spatial ones. Thus, the sampling distance can be expressed as $l_s=\left\langle v \right\rangle/f_s$. Detailed multiscale statistical analyses of Modane turbulent velocity signals have been previously provided in~\cite{Chevillard2012, Kahalerras1998, Gagne1990, Arneodo1999}. Furthermore, from previous studies the integral and Kolmogorov scales for this flow are respectively $L=2350 \, l_s$ and $\eta= 5 \, l_s$~\cite{GBelinchon2016}. Figures~\ref{fig:Struct} a), b) and c) provide respectively the evolution across scales of $\log(S^{r}_2)$, $\mathcal{S}^{r}$ and $\log(\mathcal{F}^{r}/3)$ in blue. Here and all along the paper, the natural logarithm is used and noted $\log$.

We want to point out that even if in this work the reference curves are obtained from statistical measures on real data, the used learning approach allows us to use empirical or theoretical laws to define them. Our model never sees turbulent data neither during training nor during fields generation.

We consider the optimization of our NN-Turb model according to the following four losses: 

\begin{enumerate}
\item the mean squared error between the reference $\log(S^{r}_2(l))$, $\mathcal{S}^{r}(l)$ and $\mathcal{F}^{r}(l)$, and $\log(S_2(l))$, $\mathcal{S}(l)$ and $\mathcal{F}(l)$ of the generated field $u(x)$.
\begin{eqnarray}
\mathcal{L}_{S_2} &=& \mathbb{E}_{l} \left[ \Bigl( \log(S^{r}_2(l)) - \log(S_2(l)) \Bigr)^{2} \right]  \label{eq:loss2}\\
\mathcal{L}_{\mathcal{S}} &=& \mathbb{E}_{l} \left[ \Bigl( \mathcal{S}^{r}(l) - \mathcal{S}(l) \Bigr)^{2} \right] \\
\mathcal{L}_{\mathcal{F}} &=& \mathbb{E}_{l} \left[ \Bigl( \mathcal{F}^{r}(l) - \mathcal{F}(l) \Bigr)^{2}\right]
\end{eqnarray}

\noindent where $\mathbb{E}_l$ is the expected value operator. The logarithm used only in eq.(\ref{eq:loss2}) allows to homogenize across scales the magnitude of the standard deviation of $S_2$ and, more importantly, to highlight the differences across scales of $S_2$ for scales in the dissipative and inertial ranges. This is an advantage for training the network. The logarithm is not used on the skewness because this statistic can take negative values, and it is not used on the flatness because the logarithm produces the inverse effect on this statistic: it reduces the differences of $\mathcal{F}$ across scales for scales in the dissipative and inertial ranges.

\item a regularization loss $\mathcal{L}_{C}= 1 - r(u^{\prime}(x), w(x))$ where $r$ is the cross-correlation function between $u^{\prime}(x)$, which is the centered and standardized version of $u(x)$, and $w(x)$ which is the Gaussian noise used as input. This loss aims at increasing the correlation between these two time-series.
\end{enumerate}

The optimization criterion $\mathcal{L}$ is a weighted sum of these losses: 

\begin{equation}
\mathcal{L}= \alpha \cdot \left( \mathcal{L}_{S_2} +  \mathcal{L}_{\mathcal{S}} + \mathcal{L}_{\mathcal{F}} \right) + \beta  \cdot \mathcal{L}_{C}
\end{equation}

\noindent We set empirically the weights of the three first losses to $\alpha= \alpha_{S_2} = \alpha_{\mathcal{S}} = \alpha_{\mathcal{F}} =1$ and the weight of the regularization loss to $\beta=0.1$. These weights were chosen by grid searching with a grid step of $0.1$ for $\beta \in \left[0,1\right)$ and with $\alpha_{\mathcal{S}}\in \left\lbrace 0.5,1 \right\rbrace$, $\alpha_{\mathcal{F}}\in \left\lbrace 0.5,1 \right\rbrace$ and fixed $\alpha_{S_2}=1$. We decided to impose $\alpha_{S_2}\geq\alpha_{\mathcal{S}}$ and $\alpha_{S_2}\geq\alpha_{\mathcal{F}}$ since $S_2$ directly intervenes in the definition of $\mathcal{S}$ and $\mathcal{F}$. We also imposed $\alpha_{\mathcal{S}}> \beta$ and $\alpha_{\mathcal{F}} > \beta$. Changes in $\alpha_{\mathcal{S}}$ and $\alpha_{\mathcal{F}}$ impact very slightly the results, on the other hand the value of $\beta$ was important for converging to an adequate field.

Consequently, our optimization approach completely grounds on the Kolmogorov theories as well as on previous descriptions of turbulent velocity fields. The three mean squared error losses impose the desired $2/3$ and $4/5$ laws of Kolmogorov, \textit{i.e.} the distribution and cascade of energy across scales, as well as intermittency. The regularization loss is used to impose stationarity at large scales as desired for turbulent velocity.

Using Pytorch, our learning setup relies on Adam optimizer with a learning rate of 2e-3 for the first 100 epochs, 1e-3 for epochs between 100 and 1000 and 5e-4 for remaining epochs up to epoch 2000. The open source code is available at \href{https://github.com/cgranerob/NN-Turb}{https://github.com/cgranerob/NN-Turb}.

\section{Results and discussion}
\label{sec:results}

In this section, we study the process $u(x)$ generated by our NN-Turb model $\Psi$. For this purpose, we generate $256$ realizations of $u(x)$, each one of size $N=2^{15}$ samples. To avoid border effects due to convolutions, we first generate realizations of size $N+N_b$ samples, where $N_b$ is the number of samples impacted by border effects. In our case of study $N_b=2^{13}$. Then, we only consider the $N$ samples far from the borders to define $u(x)$. We analyse the second order structure function, skewness and flatness of $u(x)$ across scales for scales going from the dissipative domain to the integral one. We perform the same analysis on Modane experimental turbulent velocity for comparison. To compare NN-Turb to current state of the art models, we also study a regularized Multifractal Random Walk, $z(x)$,~\cite{Robert2008} parameterized to simulate Modane experimental turbulent velocity: the small and large scales of regularization are $\epsilon=1$ and $R=e^{8}$ respectively, and the long-range dependence and intermittency parameters are respectively $c_1=1/3$ and $c_2=0.025$. The used large scale regularization function is Gaussian. A single realization of size $2^{23}$ was generated and sliced to provide $256$ realizations of size $N$.

\begin{figure}[!htb]
\centering
\includegraphics[width=0.6\textwidth]{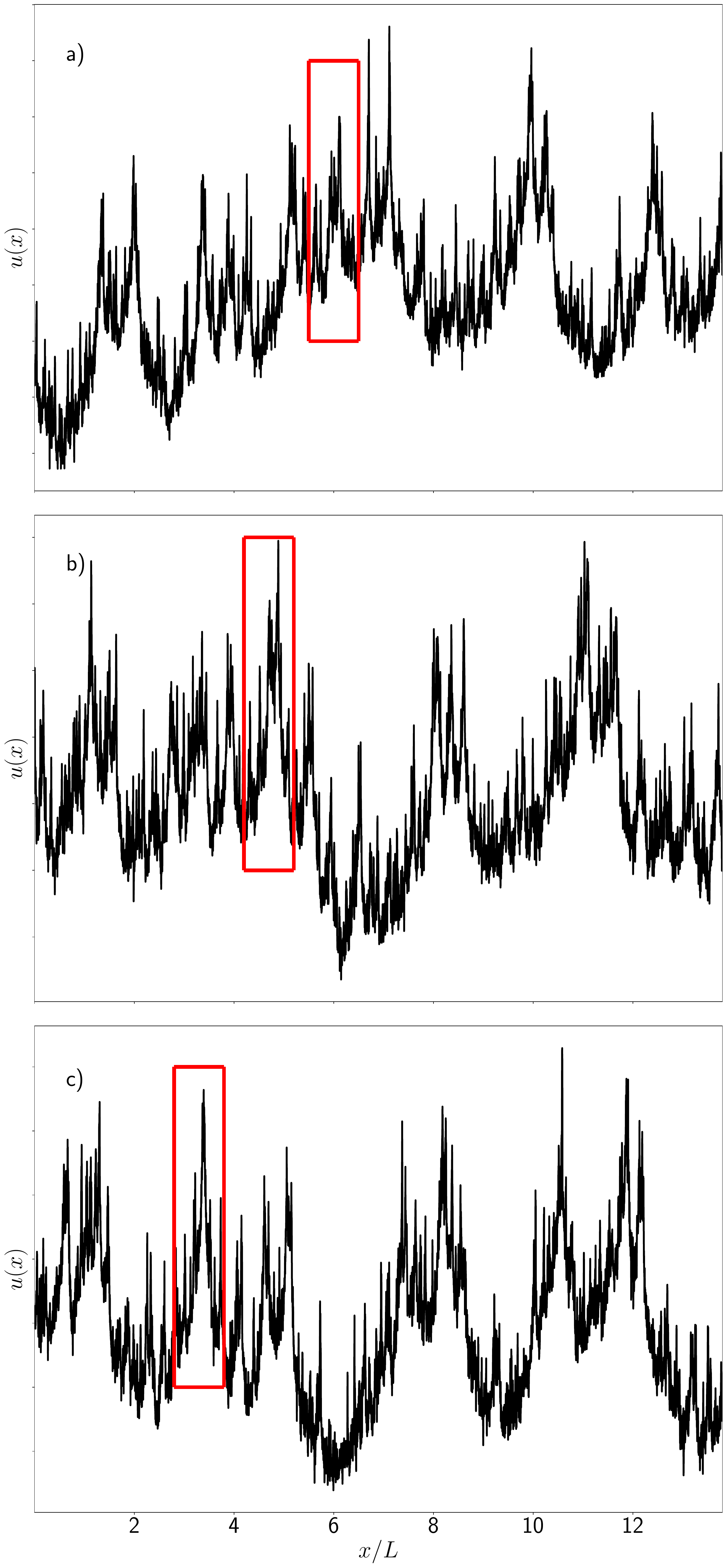}
\caption{Illustration of three realizations of process $u(x)$ generated with NN-Turb in function of the spatial variable $x/L$. The red box corresponds to the length of an integral scale $L$.}
\label{fig:Sig}
\end{figure}

Figure~\ref{fig:Sig} illustrates three realizations of $u(x)$ among the $256$ generated. We observe dynamics at very different scales, from very small scales of the order of the sampling distance $l_s$ to scales of the order of the integral scale $L$ of the process (to facilitate visualization a red box of width $L$ is shown).

\begin{figure}[!htb]
\centering
\includegraphics[width=0.6\textwidth]{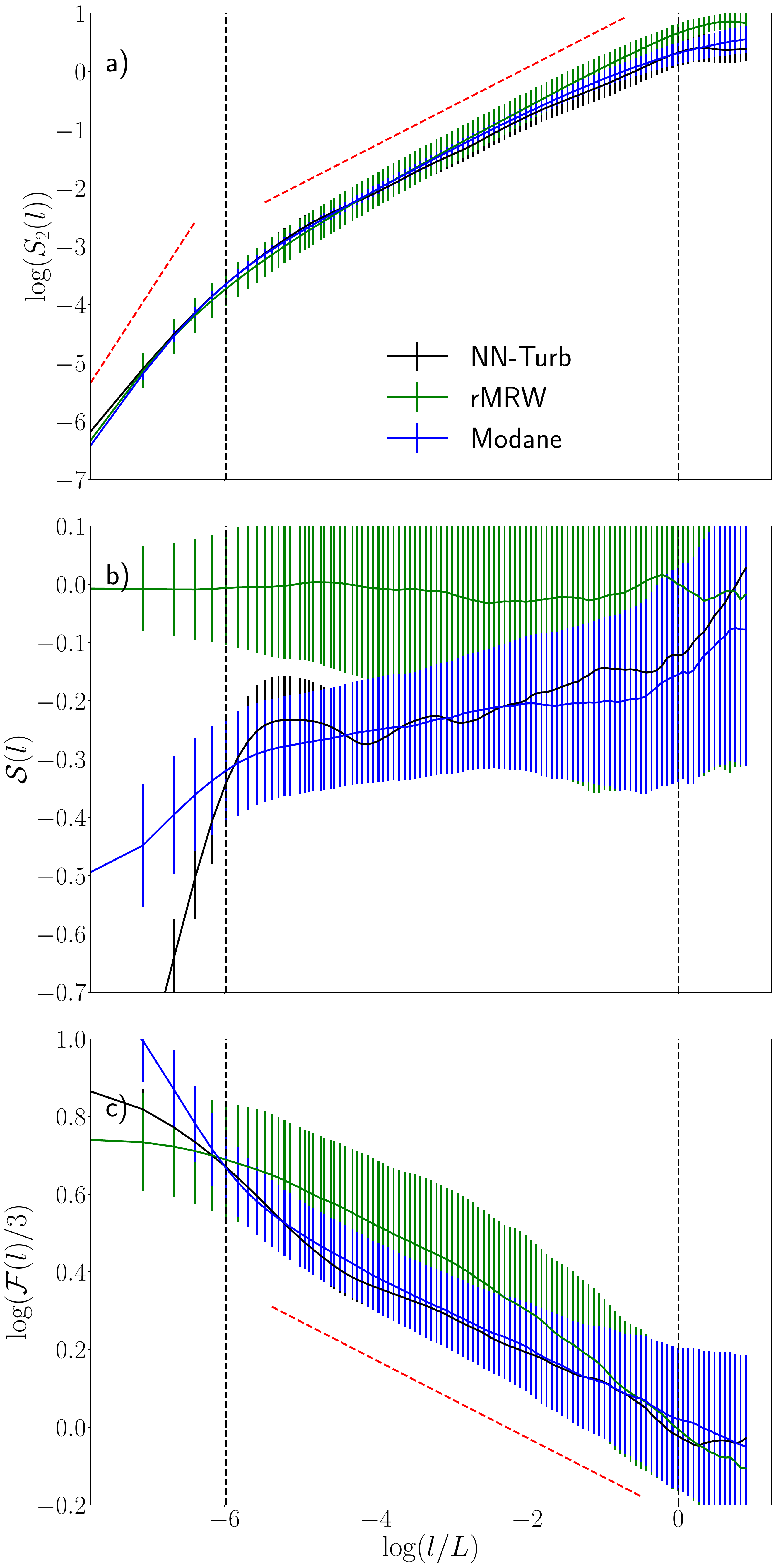}
\caption{a) Logarithm of the second order structure function $\log(S_2(l))$, b) skewness $\mathcal{S}(l)$ and c) logarithm of the flatness $\log(\mathcal{F}(l)/3)$  in function of the logarithm of the scale of analysis $\log(l/L)$ for the NN-Turb generated field (black), Modane (blue) and rMRW (green). Curves represent the mean value and errorbars the standard deviations calculated on 256 realizations. Red dashed lines in a) have a slope $2$ in the dissipative domain and $2/3$ in the inertial one describing respectively the behaviors of the Batchelor model~\cite{Batchelor1951} and the $2/3$ Kolmogorov law. Red dashed line in c) has a slope $-0.1$ previously described for the $\log(\mathcal{F}(l)/3)$ in the inertial domain~\cite{Chevillard2012}. The vertical black dashed lines correspond to the Kolmogorov and integral scales, $\eta$ and $L$ of Modane.}
\label{fig:Struct}
\end{figure}

Figure~\ref{fig:Struct} a), b) and c) show respectively $\log(S_2(l))$, $\mathcal{S}(l)$ and $\log(\mathcal{F}(l)/3)$ of the NN-Turb (black), Modane (blue) and rMRW (green) fields in function of $\log(l/L)$. The vertical dashed black lines indicate the integral and Kolmogorov scales as obtained for Modane turbulent data in previous studies~\cite{GBelinchon2016}. Thus, we observe different behaviors of the studied statistics depending on the domain of scales.

In figure~\ref{fig:Struct} a) we observe, independently of the process, a plateau of $\log(S_2(l))$ for scales $l$ larger than the integral scale $L$ as expected for turbulent velocity. These are the more energetic scales. Then, when the scale decreases through the inertial region, $\log(S_2(l))$ also decreases following the $2/3$ law of Kolmogorov. In the dissipative domain, $\log(S_2(l))$ decreases faster with the scale than in the inertial one, and its slope is close to $2$ as described by the Batchelor model~\cite{Batchelor1951}. This behavior in the dissipative domain is a signature of the smoothness of the field at these scales. While the NN-Turb field perfectly matches the Modane behavior within the errorbars for any scale of analysis, the rMRW field matches the Modane behavior in the dissipative and inertial ranges but the impact of the large scales is more localized and does not propagate through the inertial range. This leads, in the case of the rMRW process, to a linear behavior with a slope of $2/3$ remaining at larger scales compared to Modane.

Figure~\ref{fig:Struct} b) illustrates that the generated field $u(x)$ is negative skewed as requested by the Kolmogorov $4/5$ law. Moreover, due to intermittency effects in the dissipative domain, the skewness decreases whith the scale for scales $l<\eta$. However, this decrease is much more steep than the one from Modane experiment. On the other hand, the increments of the rMRW field present, by construction, null skewness indicating that their pdfs are symmetrical.

In figure~\ref{fig:Struct} c), independently of the process, $\log(\mathcal{F}(l)/3)$ goes from zero in the integral domain of scales to larger values when the scale decreases. This is a signature of intermittency in both the inertial and dissipative domains~\cite{Chevillard2012, Kolmogorov1962, Anselmet1984}. Moreover, the flatness of Modane and NN-Turb shows a linear behavior in the inertial domain with slope $-0.1$ (red dashed line). This behavior is representative of homogeneous and isotropic turbulence~\cite{Chevillard2012}. Furthermore, for these two processes, the intermittency is stronger in the dissipative domain and so the increase of flatness when the scale decreases becomes steeper~\cite{Chevillard2005}. The flatness of the rMRW process behaves otherwise and does not capture the steeper increase in the disipative domain. The slope $-0.1$ in the inertial range is reproduced but in a smaller domain of scales.

Our stochastic field recovers the correct behavior of $S_2(l)$ in all the domains of scales. In addition, it recovers the good behavior of skewness and flatness in the integral and inertial domains. However, in the dissipative domain the decrease of skewness is to steep and the increase of flatness is not enough compared to Modane statistics. This illustrates the complexity of correctly recovering high-order statistics in this region.

\begin{figure}[!htb]
\centering
\includegraphics[width=0.7\textwidth]{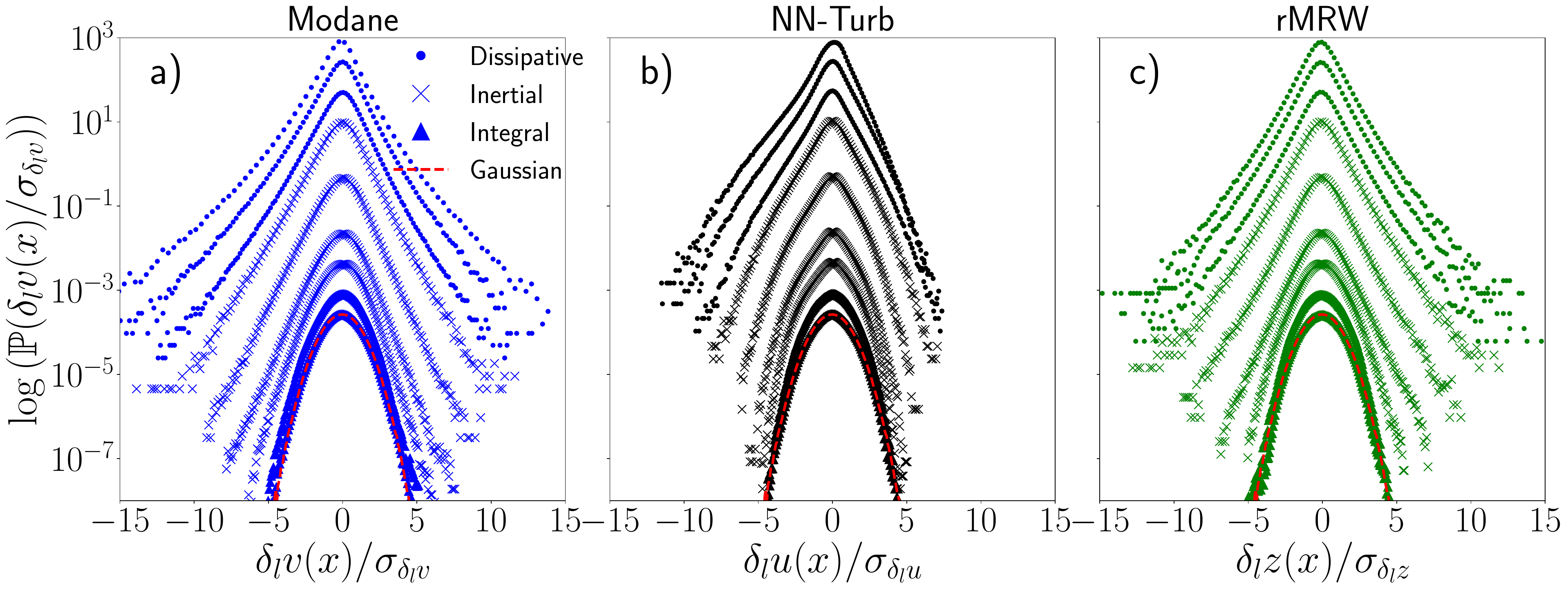}
\caption{Logarithm of the probability density function of the centered and standardized increments of a) the Modane turbulent velocity signal, b) the NN-Turb field and c) the rMRW field, in function of the values of the standardized increments. The illustrated increments are those with $l = [2,4,8,16,64,256,1024,4096,10000] \, l_s$. The integral scale of the flow is $L=2350 \, l_s$. The red dashed lines correpond to the logarithm of a Gaussian probability density function with zero mean and unit variance.}
\label{fig:pdf}
\end{figure}

Finally, figure~\ref{fig:pdf} shows the logarithm of the pdf of the standardized increments of a) the Modane turbulent velocity signal ($\delta_l v(x))/\sigma_{\delta_lv}$), b) the NN-Turb field ($\delta_l u(x))/\sigma_{\delta_lu}$) and c) the rMRW field ($\delta_l z(x))/\sigma_{\delta_lz}$), for different scales $l$.
Scales from the dissipative domain to the integral one are considered. For Modane and NN-Turb processes we observe an evolution from non-Gaussian pdfs at small scales: long-tailed and asymmetric, to close to Gaussian when approaching the integral domain (a Gaussian pdf with zero mean and unit variance is illustrated in red dashed line). So, the generated field $u(x)$ presents intermittency and becomes Gaussian at large scales. However, the evolution of the pdf of the increments of $u(x)$ is not exactly the one expected for a turbulent field~\cite{Chevillard2012}: the extreme values at small scales are only partially recovered while the asymmetry of the distribution of the increments at these scales is overvalued.
On the other hand, the rMRW process do not present asymmetries in the pdfs of its increments but better recover the broad shape of the pdfs and both the positive and negative extreme events.

\section{Conclusions}
\label{sec:conc}

We propose a fully-convolutional NN model, NN-Turb, to generate a 1-di\-men\-sio\-nal field with some turbulent velocity statistics. Our stochastic model takes as input a Gaussian white noise and perform a double operation on it: 1) it introduces the desired structure of dependencies and 2) it deformates the Gaussian pdf of the input to a long-tailed and skewed one.
Very importantly, our model only needs the aimed evolution across scales of the second order structure function, skewness and flatness for learning, and so, it does not require turbulent data.

The generated 1-dimensional field $u(x)$ correctly recovers the $2/3$ and $4/5$ laws of Kolmogorov as well as the flatness behavior in the inertial domain described in~\cite{Chevillard2005,Chevillard2012}. From this perspective, $u(x)$ models adequately the energy distribution, energy cascade and intermittency of turbulence while remaining close to Gaussian at large scales.
However, we also illustrated that the pdfs of the increments of the generated fields do not necessarily match the expected behavior of turbulent ones. So, our stochastic field can reproduce the statistical behavior of the second, third and fourth order structure functions of turbulent velocity without matching the exact pdf deformation across scales. 

Three main future perspectives are considered. 
First, the application of the proposed learning approach for generating 2D images of homogeneous and isotropic turbulent velocity. 
Second, the empirical definition of the structure functions $S_p(l)$ according to a limited number of parameters $c_m$. This will allow us to completely avoid the use of experimental data. More interestingly this will allow us to introduce these parameters $c_m$ as inputs of our model, and so, to generate different types of processes with diverse multifractal properties. Thus, we aim to generalize this NN optimization approach, which does not need data, to other kind of non-linear physical systems. 
Finally, the definition of a learning setup in which we don't impose the evolution across scales of some structure functions but the evolution of the pdfs of the increments directly.

\section*{Acknowledgment}

The author wish to thank St\'ephane G. Roux for providing the code for generating rMRW processes. This work was supported by the French National Research Agency (ANR-21-CE46-0011-01), within the program ”Appel \`a projets g\'en\'erique 2021”.

\clearpage

\appendix

\section{NN-Turb architecture}
\label{appendix:NN}

\begin{figure}[!htb]
\centering
\includegraphics[width=1\textwidth]{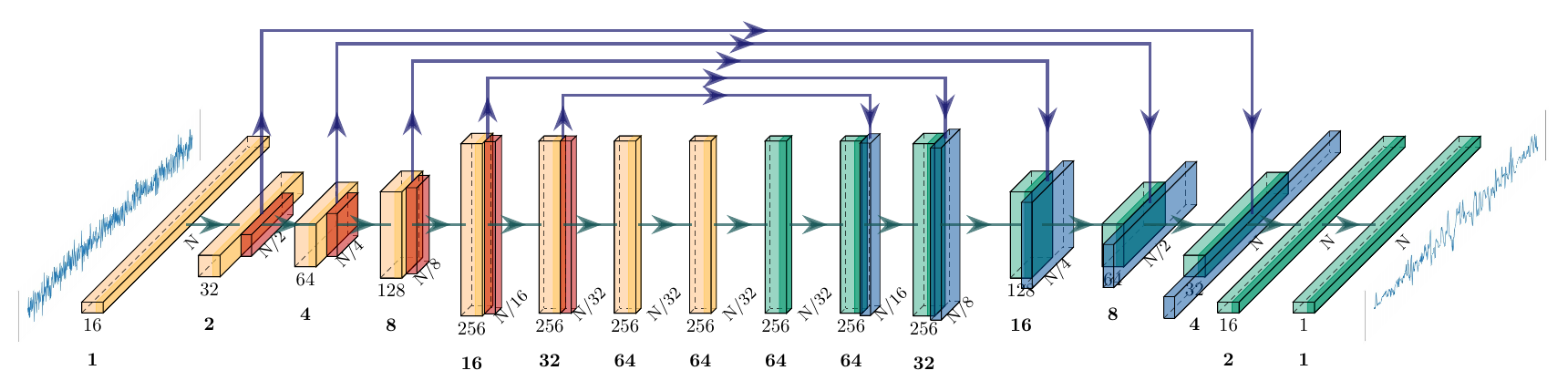}
\caption{Fully convolutional U-net architecture of NN-Turb model. The size of the input Gaussian white noise is $N$. Yellow blocks correspond to convolutional layers followed by batch normalization and ReLU activation function. Red layers correspond to average pooling of factor $2$ in the spatial dimension. Green blocks correspond to convolutional transpose layers followed by batch normalization and ReLU activation. Blue layers correspond to upscaling of factor $2$ in the spatial dimension. The kernel size of the convolutional layers is indicated in bold. The number of channels is also indicated for each block. Blue arrows indicate additive long-skip connections between the encoder and the decoder branches of the U-net.}
\label{fig:Archi}
\end{figure}

Figure~\ref{fig:Archi} illustrates the architecture of NN-Turb. It is a fully convolutional U-net architecture grounding on multi-scale decomposition to modify the Gaussian white noise used as input. 
Thus, it is composed of an encoder, and a decoder that are connected by a convolutional bridge. Because of the symmetry of the architecture, both the Gaussian white noise used as input and the output field have the same size, \textit{i.e.} the same number $N$ of samples.

The encoder blocks are the combination of 1{\sc d} convolution layer with batch normalization, non-linear ReLU activation function and average pooling. 
The decoder blocks are the combination of 1{\sc d} convolution transpose layer with batch normalization, non-linear ReLU activation and upsampling layer. 
A bridge with a 1{\sc d} convolution layer with batch normalization and non-linear ReLU activation followed by a 1{\sc d} convolution transpose layer with batch normalization and non-linear ReLU activation is used to connect the encoder and the decoder of the U-net. 
The number of channels evolves as follows: $1 \rightarrow 16 \rightarrow 32 \rightarrow 64 \rightarrow 128 \rightarrow 256  \rightarrow 128 \rightarrow 64 \rightarrow 32 \rightarrow 16 \rightarrow 1$) and the kernel sizes are $[1,2,4,8,16,32,64]$. Furthermore, we introduced additive long-skip connections between the encoder and the decoder layers~\cite{Ronneberger2015}.

We want to point out that in both the encoder and the decoder the kernel sizes of the different layers increase exponentially in order to rapidly and completely sample the dissipative, inertial and integral domains of turbulence. This is specially important to recover the expected multi-scale behavior of each domain.
Furthermore, the average pooling and upscaling layers also facilitate to process the whole domain of scales of interest without increasing dramatically the computational cost of learning. 

The additive long-skip connections have been introduced to facilitate the learning~\cite{Ronneberger2015}. Moreover, the longest skip connection, the one connecting the second and next-to-last layers, appeared as crutial to impose Gaussianity at large scales.

Furthermore, we decided to generate the smallest available velocity increment $\delta_{l_s}u(x)$ instead of directly the turbulent velocity $u(x)$ since the long-range dependencies of the velocity could complicate the learning process. The range of dependencies of the velocity increments are shorter.

\clearpage

\bibliographystyle{elsarticle-num} 
\bibliography{THEBIBLIO}

\end{document}